# Kinetics of Intrinsic Stress in Nanocrystalline Films

Enrique Vasco[1*], María J. Ramírez-Peral[1,2,4], Alfredo Jacas-Rodríguez[1] and Celia Polop[2,3,4]

[1]*Instituto de Ciencia de Materiales de Madrid, CSIC, Sor Juana Inés de la Cruz 3, 28049 Madrid, Spain*
[2]*Departamento de Física de la Materia Condensada, Universidad Autónoma de Madrid, 28049 Madrid, Spain*
[3]*Condensed Matter Physics Center (IFIMAC), Universidad Autónoma de Madrid, 28049 Madrid, Spain*
[4]*Instituto Nicolás Cabrera, Univ. Autónoma de Madrid, Spain*

Conventional polycrystalline materials acquire high levels of intrinsic mechanical stress (ranging from MPa to a few GPa) during preparation and use, but this stress decays quickly (~minutes) to small residual values (~kPa) under standard resting conditions. Nanocrystalline materials reach similar or even higher levels of intrinsic stress, but surprisingly retain a significant portion of this stress over much longer time scales (~hours). This behavior directly contradicts current theoretical models that predict stress relaxation through diffusive currents. Diffusive currents, which flow mainly on the surfaces of grains, are expected to produce faster relaxation kinetics when the stress to be released is higher. In this work, we study the kinetics of intrinsic stress relaxation in nanocrystalline films and identify the limitations of this process as a preliminary step towards designing a strategy for high-stress stabilization in nanostructured systems.

[*]Corresponding author: enrique.vasco@csic.es

The origin and dynamics of intrinsic compression in polycrystalline films and coatings has been intensively studied for decades [1], yet some experimentally verified features of this phenomenon still have no convincing theoretical explanation. This work proposes a robust theoretical model to explain the unusually long-time scales of intrinsic stress relaxation in nanocrystalline films.

In principle, intrinsic compression is correlated with the presence of grain boundaries, which alter the surface kinetics of these systems under conditions of high atomic transport [2,3,4]. The compression strength rises as the density of grain boundaries (~1/grain size) increases [5,6], so finer-grained films exhibit higher compressions. Once the condition of high atomic transport is removed, the intrinsic compression relaxes via currents of adatoms driven by gradients of stress [7,8] and/or strain energy [9,10]. The currents flow from regions under compression towards regions under traction through relaxed zones. The strength of these currents, which determines the relaxation kinetics, depends on the magnitude of the stress to be released. For example, the transport of adatoms from positions with a strain energy greater than the energy $k_B T$ of the "thermal bath" of the system would be biased rather than a random walk. On the other hand, regions that are only slightly strained (having strain energy much lower than the diffusion barrier) retain a portion of their intrinsic stress (residual stress) for much longer. This explanation of how stress relaxation kinetics relate to the magnitude of stress has been experimentally corroborated for micron and sub-micron grains [11], and also mesoscopically modelled [3,10].

However, several groups [5,12] have reported surprising behavior of the relaxation kinetics for grain sizes in the nanometer range (i.e., in nanocrystalline films). As might be envisioned, nanocrystalline films exhibit huge intrinsic compressions; however, their relaxation kinetics are much slower than expected, and their stress levels are retained longer than the lower stress levels observed in films with larger grains [5]. This behavior was investigated by Yu *et al.* [12] in films with a bimodal grain size distribution. They interpreted the exponential decay of intrinsic stress as two coexisting relaxation processes: a fast one (~$10^2$ s) associated with large grains (submicron sizes) and a slow one (~$10^4$ s) ascribed to the growth of small grains (with sizes ~30-60 nm) [12]. We can reinterpret grain growth as a shortcut to stress relaxation, as discussed below. In this work, we address this unusual property of nanocrystalline films, which we believe can be used in stress/strain engineering as a strategy to stabilize high stress levels in a nanostructured system comprising layers, coatings, and surfaces.

Equation 1 describes the morphology evolution of a film from the local advance rate of its surface $h(\vec{r}, t)$ according to the continuity equation:

$$\frac{1}{\Omega}\partial_t h = \vec{\nabla} \cdot \vec{j}_s \tag{1a}$$

where the net surface diffusion current $\vec{j}_s$ depends on the surface chemical potential $\mu$ and the atomic drift transport $\Phi$ as predicted by Fick's first law:

$$\vec{j}_s = -\Phi \vec{\nabla}_s \mu \tag{1b}$$

with $\Phi = D_s n_1 / k_B T$ [2] and $\mu$ containing the following contributions:

$$\mu = \Omega \left[ \gamma_s \nabla_s^2 h + N \left( \sigma + \nu \frac{\sigma^2}{2M} + \cdots \right) \right] \tag{1c},$$

The first term is the curvature of the surface $\kappa \propto \nabla_s^2 h$, which provides an estimate of the local density of dangling bonds. The second term is the local stress field $\sigma$, including first (i.e., strength), second (strain energy $\varepsilon = \sigma^2/2M$), and higher powers as needed. In Eqs. 1a-b, $D_s$ is the surface diffusion coefficient, $\Omega$ is the atomic volume, $n_1$ is the steady density of diffusing adatoms (monomers) far from the grain boundary (GB), and $k_B T$ has its usual meaning. In Eq. 1c, $\gamma_s$ denotes the surface tension, and $M$ is the biaxial elastic modulus of the film. The parameters $N = n_{ex}/n_1$ and $\nu = \sigma/|\sigma|$ are dimensionless, describing the excess of adatoms $n_{ex}$ condensed at the edges of the GB groove and the sign of $\mu$ ascribed to the strain-energy term, respectively. Further description of these two parameters and their role is provided below.

Substituting Eq. 1c into Eq. 1b, the net surface diffusion current $\vec{j}_s$ can be divided into contributions: $\vec{j}_s = \sum_i \vec{j}_i = \vec{j}_k + \vec{j}_\sigma + \vec{j}_\varepsilon + \cdots$, where the subscripts $k$, $\sigma$, and $\varepsilon$ refer to the curvature, stress strength, and strain energy contributions, respectively.

**During deposition**, the surface diffusion currents driven by curvature $\vec{j}_k$ collect adatoms from within a **λ**-wide strip near the GB (**λ** denotes the diffusion length) and transport them to the GB where the curvature reaches a maximum. According to our model [13], this transport causes an excess of adatoms at the edges of the GB groove.

The excess then condenses, inducing overgrowth in the form of ridges with volume $V_{ridge} = \Omega n_{ex} = \Omega n_1 N$ per surface unit. As the ridge increases in volume, its curvature $\kappa \approx 8 h_{ridge}/\omega^2$ [10] also increases (where $h_{ridge} = max[\Omega n_{ex}(r)]$ and $\omega$ are the height and ridge width), inducing a Laplace compression $\sigma = 2\gamma_s \kappa \approx 16 \gamma_s h_{ridge}/\omega^2$. This compression enables the surface diffusion currents driven by stress $\vec{j}_\sigma + \vec{j}_\varepsilon$ to detach and transport adatoms back from the ridges. When a balance between $\vec{j}_k$ and $\vec{j}_\sigma + \vec{j}_\varepsilon$ is established (which implies $\vec{j}_s \approx 0 \Rightarrow \partial_t h \approx 0$), the Laplace compression reaches its experimentally observed stationary value [1].

**After deposition** (once the flux stops, $F = 0$), $n_1$ vanishes (because $n_1 \propto F/D_s$ [4]), so $\vec{j}_k$ falls sharply in the absence of mobile species. On the other hand, the currents $\vec{j}_\sigma + \vec{j}_\varepsilon$ are still present (since they depend on the parameter $N$). These currents detach and disperse the $n_{ex}$ adatoms in the ridge to reduce stress gradients. Since the intrinsic compressive stress is understood as a local excess of atoms, interatomic repulsions predominate. In this context, traction is ascribed to rarefied regions dominated by interatomic attractions, so surface diffusion currents driven by stress relaxation $\vec{j}_\sigma$ and strain-energy relief $\vec{j}_\varepsilon$ flow from dense to rarefied regions. The fact that the strain energy is independent of the stress sign makes it necessary to define the parameter $v$, in order to bias $\vec{j}_\varepsilon$ and avoid diffusion outward from the regions under traction.

For moderate stresses, e.g., with values $\leq -1$ GPa in metals (where $M \sim 10^2$ GPa), the surface diffusion currents driven by strain-energy relief are negligible (lower than 1% of the net surface current). Their contribution rises quickly for higher stresses; however, this regime is beyond the scope of our study. Consequently, we assume $\vec{j}_\sigma + \vec{j}_\varepsilon \approx \vec{j}_\sigma$ so that the balance between surface diffusion currents becomes:

$$n_{ex} \underset{J_\sigma}{\overset{J_k}{\leftrightarrows}} n_1 \qquad (2)$$

The kinetics of stress relaxation and recovery are studied in Refs. [3,10]. These works find that the stress kinetics depend on the morphological evolution of the surface, and in particular on the local height of the ridges: $\partial_t \sigma \propto -\left(\frac{1}{\omega^2}\right) \partial_t h_{ridge}$ [10]. This estimate predicts that the rate of stress relaxation decreases as the excess atoms in the ridge are redistributed over a larger surface area, implying $\partial_t h_{ridge} < 0$. By replacing Eqs. 1(a-c) in this estimate, while also assuming that the ridge height is the local maximum of the surface, such that $\partial_t h_{ridge} = \partial_t h(\vec{r} \to r_{ridge}, t)$, and then grouping the parameters in $\chi_i$, we get:

$$\partial_t |\sigma| \propto \vec{\nabla}_s \cdot (\vec{J}_k + \vec{J}_\sigma + \vec{J}_\varepsilon + \cdots) \propto -\sum_i \chi_i |\sigma|^i \qquad (3).$$

The absolute value $|\sigma|$ is used here to simplify the description of relaxation phenomena without including the stress sign convention. Eq. 3 quantifies the phenomenon that we described in the Introduction, that the relaxation kinetics (i.e., its rate) depends on the surface diffusion currents, which in turn are a function of the magnitude of the stress to be released. Roughly, $\partial_t |\sigma| \approx -\chi_1 |\sigma|$ describes an exponential decay in $\sigma$ with the relaxation time.

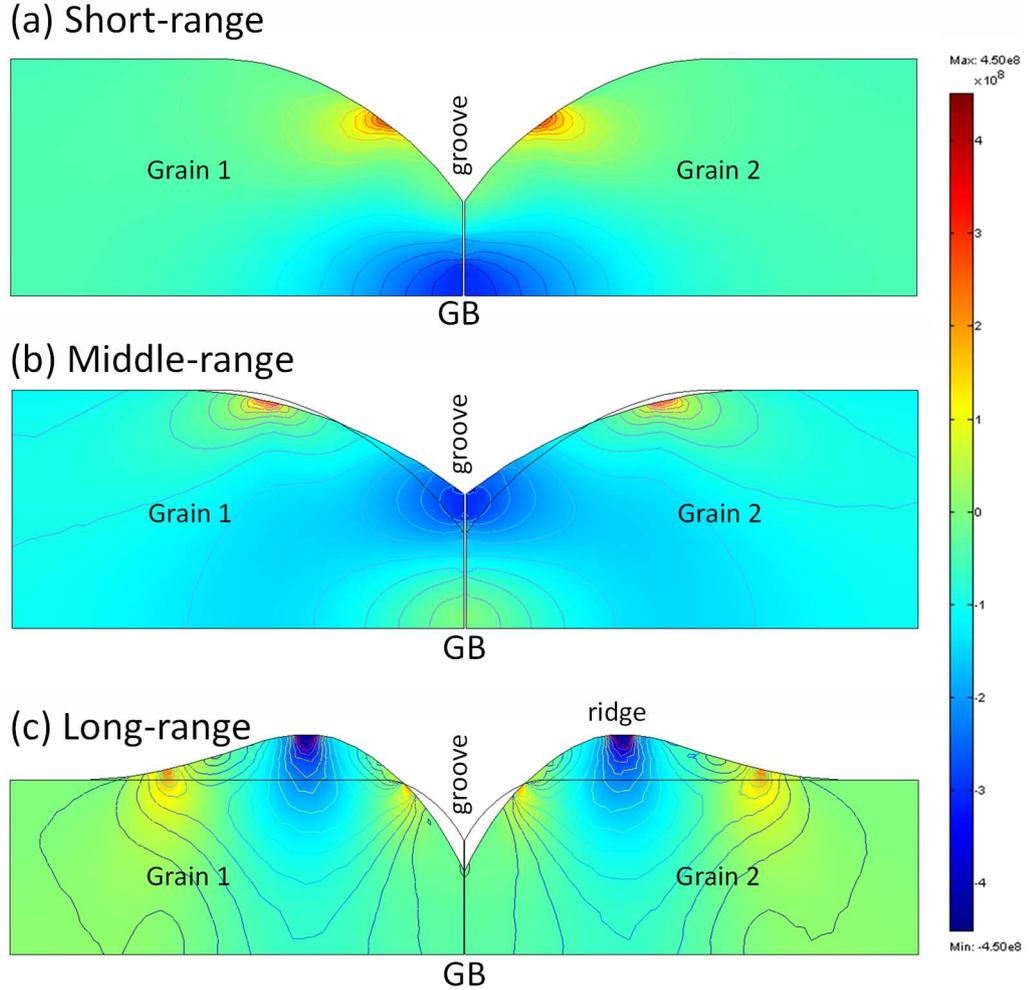

**Figure 1.** Summary of different intrinsic stress models computed by FEM [15] in the vicinity of a GB. The models are classified by the spatial range of the relaxation of the intrinsic stress field (whose strength and sign are scaled by color). The models proposed are: (a) short-range relaxation [7,8]; (b) mid-range relaxation [14,16]; and (c) long-range relaxation [10,13]. The empty curves (other than the contours) represent the equilibrium profiles.

Various models have been proposed and updated over the past two decades to address the origin of intrinsic compression in compact polycrystalline films. Rather than organizing them by the nature of the stress (i.e., kinetic or thermodynamic), here we look at models with different spatial ranges for the relaxation of the intrinsic stress field. We further assume (as discussed above) that stress fields are generated around the GBs while the grain bulk remains mostly relaxed. Figure 1 illustrates the intrinsic stress fields for relaxation mechanisms with different spatial ranges. These stress fields are calculated by finite-element modeling (FEM) using the simulation in Ref. [8], which is slightly modified to qualitatively reproduce the previous models (those shown in Figs. 1 [7], 3 [14] and 3 [13]). Details of the FEM calculation are provided in Ref. [15].

**At short range** (Fig. 1a), the model proposed by Tello et al. [7] and its variants [3] attribute intrinsic compression to the reversible insertion of adatoms inside GBs. On the other hand, the model proposed by González *et al.* [8] predicts that the intrinsic compression is the result of repulsion between bundles of misoriented grains embedded in compact media. In both models the intrinsic stress field has a short-range dipole structure along the GB profile, with compression (blue regions) below the triple-junction point and traction (yellow-red) within the groove. **In the middle range** (Fig. 1b), Thompson and coworkers [14,16] proposed the formation of GB grooves with non-equilibrium shapes due to copious fluxes and/or or diffusions towards the GBs. These non-equilibrium shapes are characterized by larger dihedral angles, which generate intrinsic compression in the vicinity of the triple-junction point, and traction near the edges of the GB groove. **At long range** (Fig. 1c), Vasco and coworkers [10,13] recently proposed that intrinsic compression is the result of the ridge-shaped buildup of adatoms at the edges of the GB groove, which occurs due to kinetic limitations in interlayer transport towards the GBs [4]. This buildup was predicted early by Mullins´ Theory [17] as a consequence of surface diffusion along the GB groove profile, where the surface chemical potential $\mu$ exhibits a gradient between the free surface (away from the boundary) and the GB triple-junction point. The ridge-shaped buildup locally modifies the surface curvature resulting in a sinusoidal capillary stress profile: slight traction within the GB groove, dominant compression at the edges, and traction beyond, while most of the grain remains relaxed.

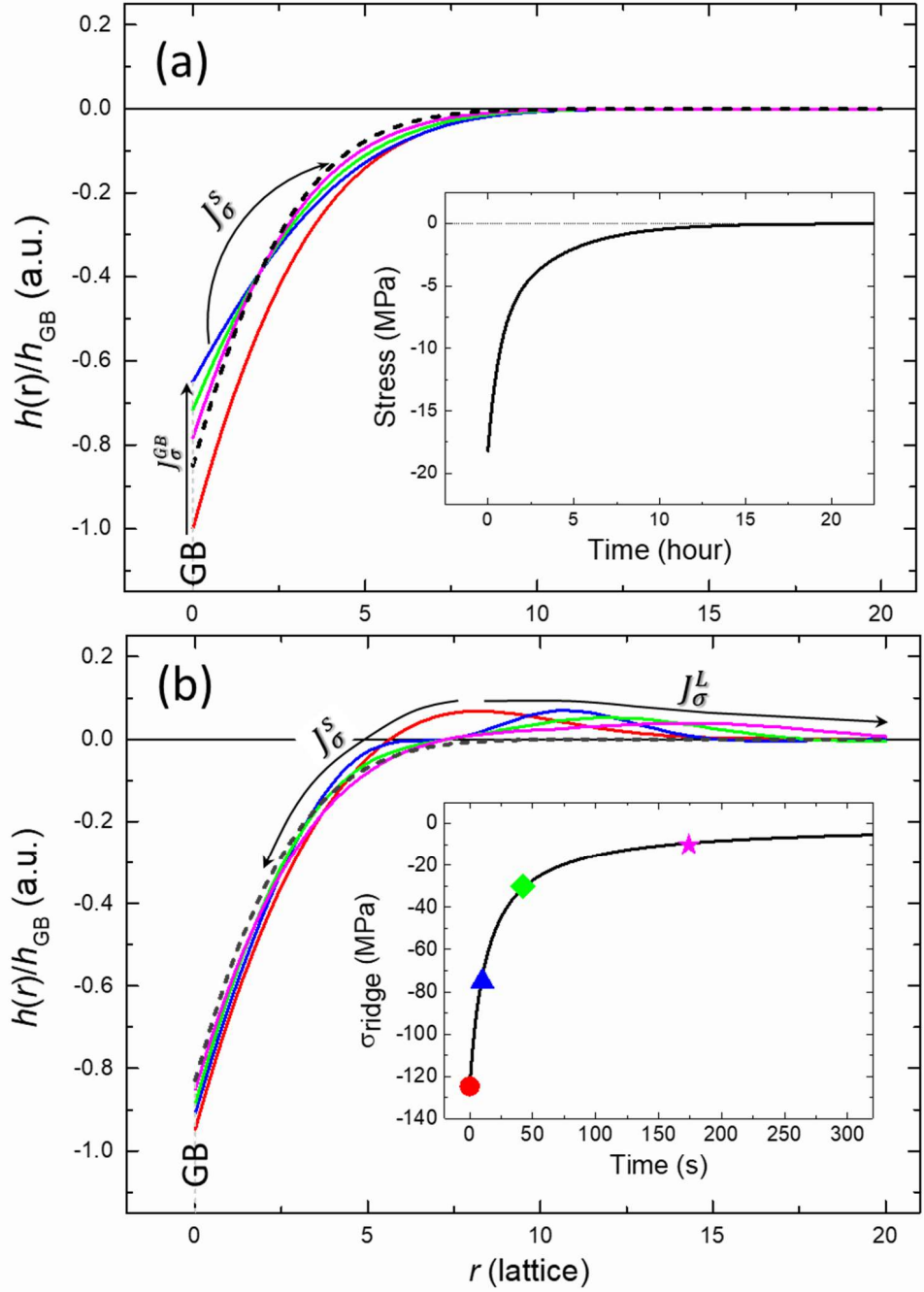

**Figure 2.** Morphology evolution of the GB profiles during (a) short/mid-range and (b) long-range stress relaxation, together with the corresponding kinetics of stress relaxation (insets). Except for the inset in (a), which is taken from Fig. 2 in Ref. [8] and shown in linear-linear plot, the data are newly calculated for this work. (Main graphs) The solid red curves show the morphologies resulting from deposition (before relaxation—initial state). The solid, colored curves show intermediate morphologies at arbitrary relaxation times. The dashed black curves depict the equilibrium morphologies (asymptotic final state). The progression is solid red->blue->green->magenta->dashed black. (Insets) The kinetics of stress relaxation (in b, the points follow the color convention of the curves). Labelled arrows indicate the direction of the diffusion currents described in the text.

The morphology evolution resulting from stress relaxation can be computed by applying Eqs. 1 (a-c) to the FEM results shown in Fig. 1. The evolving profiles are shown in Figures 2a and 2b for the short-range and long-range relaxation models, respectively. The figure insets give information on the corresponding relaxation kinetics, i.e., the $\sigma$ decay over time (according to Eq. 3). The kinetic data of the short-range case (Fig. 2a inset) are taken from Ref. [8], while the kinetics of the long-range relaxation are calculated for the first time here [18].

The calculation for the short-range relaxation models (Fig. 2a) was artificially divided into two consecutive steps: a slow diffusion $\vec{j}_\sigma^{GB}$ (over several hours) along the GB of all atoms expelled from the compressed bulk toward the surface, and a fast short-range surface diffusion $\vec{j}_\sigma^S$ toward the edges of the GB groove. Comparing the characteristic time scales of the two processes [19] indicates that lateral reconstruction of the GB groove (through $\vec{j}_\sigma^S$) occurs simultaneously as the atoms are expelled (through $\vec{j}_\sigma^{GB}$) [8]; hence the two processes are not formally separable. However, the two-step calculation has the following advantages: it is not expected to generate much error when the shape of the GB groove is close to the equilibrium profile (dashed black curves), it is simple to calculate, and it is more intuitive for the reader. Furthermore, we note that the intrinsic stress field of the intermediate state (blue profile in Fig. 2a) recreates the profile of the mid-range relaxation models. This implies that the two-step model calculated in Fig. 2a corresponds to the relaxation of both stress fields shown in Fig. 1a and 1b.

On the other hand, the long-range relaxation model (Fig. 2b) describes a more complex evolution. The ridge at the edge of a GB groove dissociates into two lobes that disperse and redistribute the mass in opposite directions. A small, short-range current $\vec{j}_\sigma^S$ flows towards the GB triple-junction, but most of the adatoms diffuse towards the relaxed top of the grains (long-range current $\vec{j}_\sigma^L$). The fact that the diffusion induced by stress relaxation occurs preferentially towards the top of the grains is due, as previously discussed in [10], to the increasing density of step-edge barriers [4] as we approach the triple-junction point where slope constriction defines a maximum density. From Eqs. 1(a-c), we can estimate $\vec{j}_\sigma^L + \vec{j}_\sigma^S \approx \vec{j}_\sigma^L = \frac{D_s}{k_B T} n_{ex}(-\Omega \vec{\nabla}_s \sigma)$, where the parentheses enclose the driving force of $\vec{j}_\sigma^L$ and the outer terms correspond to its kinetic parameters.

A comparison between the two types of evolution (Figs. 2a and 2b) reveals some interesting findings. While short-range relaxation causes a slow *filling* of the GB groove by diffusion along GBs and groove restructuring, long-range relaxation hardly affects the profile of the GB groove. Instead, the adatoms piled up at the edges of the GB grooves diffuse in a matter of minutes towards the relaxed top of the grains. The relaxation of intrinsic stress by this mechanism takes place on time scales of a few minutes (inset in Fig. 2b and Ref. [5]), much shorter than the time scale required by diffusion along GBs (several hours—inset in Fig. 2a). This time difference is mainly due to the fact that the diffusion coefficient along the GB is several orders of magnitude lower than the surface diffusion coefficient [19]. This causes the difference between the magnitudes of the intrinsic stresses to relax (here, -18 MPa in Fig. 2a inset and -123 MPa in Fig. 2b inset, which were taken from Ref. [8] and Ref. [4], respectively]) plays a minor role in this effect. Note that during the short time scale the GB grooves do not change shape significantly [20], further indicating that diffusion of previously inserted adatoms along the GBs does not contribute significantly to this process. Another notable difference is the fact that short-range relaxations occur in an isolated manner, without affecting neighboring GBs. In contrast, long-range relaxations allow potential interactions between GBs with spacing shorter than the diffusion length **λ**. This is the case for nanocrystalline films, as discussed below.

Once the $n_{ex}$ adatoms detach from the ridges and diffuse through $\vec{j}_\sigma^L$ towards the relaxed top of the grains, they become mobile species that (according to Eq. 2) contribute to increase $n_1$. A small fraction of these adatoms diffuses back towards the GB via the surface curvature-biased current $\vec{j}_k$, but most of them decay via secondary nucleation (i.e., diffusing adatom meets diffusing adatom [4]). Those diffusing back towards the GB by $\vec{j}_k$ slightly delay the stress relaxation kinetics around an isolated GB (as shown in Fig. 2b inset), since later they have to be redispersed again by $\vec{j}_\sigma^L$. The balance between $\vec{j}_\sigma^L$ and $\vec{j}_k$ (with $J_\sigma^L \geq J_k$ for relaxation to occur) is ruled by the length scales at which the surface diffusion currents operate. While $\vec{j}_\sigma^L$ disperses the adatoms as far as required to overcome local stress gradients, $\vec{j}_k$ collects them within a **λ**-wide strip near the GB.

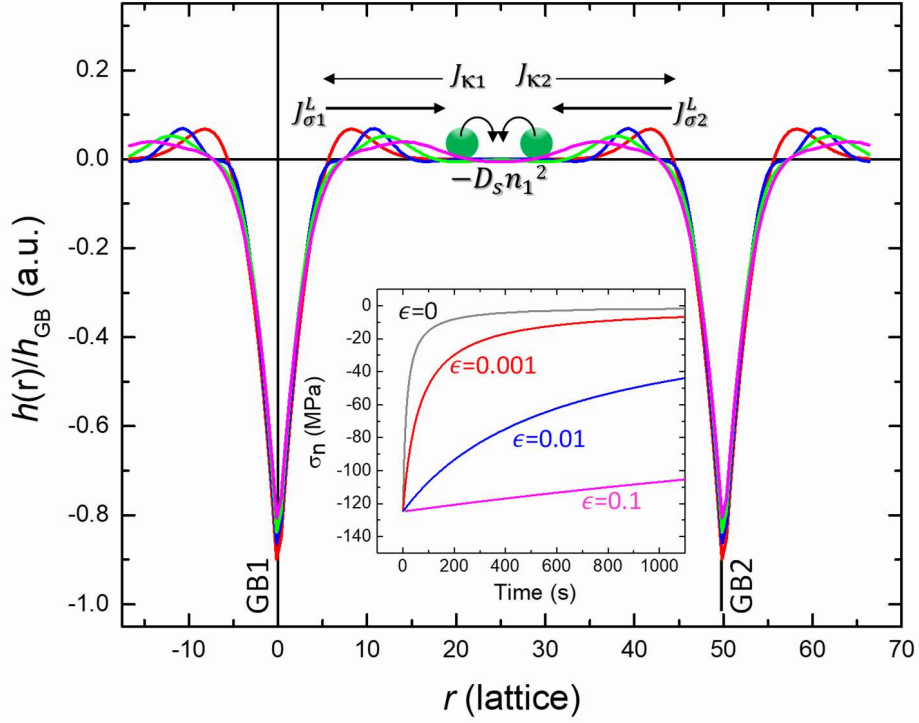

**Figure 3.** (Main graph) Balance between long-range diffusion currents on the surface of a $\xi = 50$ nm-sized grain giving rise to the adatom exchange between GBs and the secondary nucleation events. (Inset) Kinetics of intrinsic stress relaxation for different exchange coefficients $\epsilon$ (as $\epsilon$ increases, the nanocrystalline behavior is accentuated [18]).

For small grains (i.e., nanocrystals) with sizes $\xi$ lower than $\lambda$, the areas (diffusion fields) where the adatoms are dispersed by $\vec{j}_\sigma^L$ and collected by $\vec{j}_k$ towards different GBs overlap, resulting in an exchange of adatoms between GBs. Figure 3 (main graph) shows an example of adatom exchange between opposite GBs of a 50 nm-sized grain. According to Eq. 3, the relaxation kinetics of the intrinsic stress around GB1 and GB2 are given by:

$$\partial_t|\sigma_1| \propto \vec{\nabla}_s \cdot \left(-J_{\sigma_1}^L + J_{k_1}\right) - D_s n_1^2 \qquad (4a)$$

$$\partial_t|\sigma_2| \propto \vec{\nabla}_s \cdot \left(-J_{\sigma_2}^L + J_{k_2}\right) - D_s n_1^2 \qquad (4b)$$

where $J_\sigma^L$ and $J_k$ are the long-range relaxation and curvature-biased currents respectively, as discussed above and outlined in Fig. 3. The term $-D_s n_1^2$ denotes the decay in $n_1$ due to the secondary nucleation [4]. Here, it is easy to show that in the absence of a secondary nucleation (assuming the last terms in Eqs. 4(a-b) to be negligible) $\partial_t|\sigma_1| = -\partial_t|\sigma_2|$, which means that the intrinsic stress around GB1 only relaxes at the expense of generating new stress around GB2, and vice versa. The

adatom exchange between ridges homogenizes the net intrinsic stress $\sigma_n = \sum_i \sigma_i$ in the film but does not decrease it. From Eqs. 4(a-b), $\partial_t |\sigma_n| = \partial_t \sum_i \sigma_i = \partial_t |\sigma_1| + \partial_t |\sigma_2| + \cdots \propto -D_s n_1^2$, which indicates that the net intrinsic stress relaxes at a rate ruled by secondary nucleation events. The magnitude of the adatom exchange is quantified by a dimensionless exchange coefficient $\epsilon \approx (2\lambda/\xi)(J_\sigma^L/J_\kappa)$ [18], which describes the proportion of adatoms carried by $\vec{j}_\sigma^L$ that diffuse back (via $\vec{j}_\kappa$) towards less stressed GBs. The inset in Fig. 3 shows the relaxation kinetics of $\sigma_n$ for different values of $\epsilon$. A small increase in $\epsilon$ [from no exchange, $\epsilon = 0$ (polycrystalline behavior), to a moderate exchange, $\epsilon = 0.1$ (nanocrystalline behavior)] prevents the secondary nucleation from occurring in the short term, which significantly delays the stress relaxation kinetics from a few minutes to many hours. Note that this time delay is sufficient to explain the slow relaxation kinetics reported so far in nanocrystals [5,12], which suggests that heavier exchanges (with $\epsilon > 0.1$) are not required to address the current data. From a thermodynamic point of view, the exchange of adatoms between GBs in nanocrystals can be understood as a major kinetic limitation. In practical terms, these long relaxation times are limited by the grain growth rate under operating conditions (as reported in [12]), since the irreversible coarsening of nanocrystals into larger structures (with less ridge length per unit area and reduced overlap of the diffusion fields across the grain diameter) is a shortcut to enable fast relaxation of the intrinsic stress to occur in these systems.

In summary, the ridge-shaped buildup of adatoms at the edges of the GB groove, leading to intrinsic capillary compression in polycrystalline films, decays by surface diffusion once the conditions of high atomic transport are discontinued. Decay kinetics is delayed when diffusion fields coming from different GBs overlap, which preferably occurs in nanocrystals. This property can be used to retain high levels of residual mechanical stress for a long time. Such high levels have many uses: for example, to enhance the mobility of carriers in MOSFET channels, improve coating strength, improve the catalytic activity of surfaces, or as a nanostructuring tool. Tailoring the characteristic length of crystal coherence and grain growth rate to stabilize high stress levels is a novel stress/strain engineering strategy.

**ACKNOWLEDGMENTS**

This work was supported by the Ministerio de Ciencia, Innovación y Universidades (Spain) within the framework of UE M-ERA.NET 2018 program, under Project StressLIC (Spanish subprojects PCI2019-103604 and PCI2019-103594), and by the Spanish MICINN under grant nr. FIS2017-82415-R. M.J.R.P. acknowledges support by Comunidad de Madrid under contract PEJ-2019-AI/IND-14228. C.P. acknowledges financial support from the MICINN, through the "María de Maeztu" Programme for Units of Excellence in R&D (CEX2018-000805-M).

**DATA AVAILABILITY STATEMENT**

The data that support the findings of this study are available from the corresponding author upon reasonable request.

# SUPPLEMENTARY INFORMATION

## Kinetics of Intrinsic Stress in Nanocrystalline Films

Enrique Vasco[1*], María J. Ramírez-Peral[1,2], Alfredo Jacas-Rodríguez[1] and Celia Polop[2,3,4]

[1]*Instituto de Ciencia de Materiales de Madrid, CSIC, Sor Juana Inés de la Cruz 3, 28049 Madrid, Spain*

[2]*Departamento de Física de la Materia Condensada, Universidad Autónoma de Madrid, 28049 Madrid, Spain*

[3]*Condensed Matter Physics Center (IFIMAC), Universidad Autónoma de Madrid, 28049 Madrid, Spain*

[4]*Instituto Universitario de Ciencia de Materiales Nicolás Cabrera, Univ. Autónoma de Madrid, Spain*

(Dated: May. 18, 21)

**Index:**

**(A) Steady-state density of adatoms in polycrystalline thin films**

The steady-state density of adatoms far from the grain boundary (GB), $n_1^s$, defined by the condition $\partial_t(n_1 = n_1^s) = 0$, is calculated for particular scenarios from the following mean-field rate equation:

$$\partial_t n_1(t) = F + \Gamma n'_{step} - D_s n_1 n_{step} - D_s n_1^2 . \qquad (S1),$$

The first two terms correspond to increases in adatom density, originating from a deposit flux and/or detachment from surface steps under compression, respectively. The last two terms account for the decay in adatom density due to the capture of adatoms by steps and/or secondary nucleation (i.e., diffusing adatom meets diffusing adatom). $F$, $\Gamma$ and $D_s$ denote the deposition flux, detachment rate and diffusion coefficient, respectively. The spatial dependencies of the densities of adatoms $n_1(\vec{r})$ and step sites $n_{step}(\vec{r})$ are discussed elsewhere [Vasco20]. In brief, the model assumes that the steps are confined near the GBs, resulting in a local depletion of adatoms. $n'_{step}$ is a subset of $n_{step}$ that denotes the density of steps at the ridge under compression from which the adatoms detach. The scenarios to consider are:

(***During steady deposit, $F > 0$***) adatoms come mainly from the flux and are initially transported towards GBs by the curvature-biased current $\vec{j}_k$. This transport persists until $\vec{j}_k$ is compensated by the stress-biased current $\vec{j}_\sigma^L$ coming from the detachment of adatoms from ridges under Laplace compression, which transports the adatoms back towards the relaxed tops of the grains. The balance between $\vec{j}_k$ and $\vec{j}_\sigma^L$ cancels the net transport of adatoms (the second and third term in Eq. S1 cancel each other). Eq. S1 reduces to $\partial_t n_1(t) = F - D_s n_1^2$, which implies the steady-state value $n_1^s = \sqrt{F/D_s}$.

(***After deposit, $F = 0$***) adatoms are supplied from the detachment of the ridges under Laplace compression, and then transported towards the relaxed tops of the grains by $\vec{j}_\sigma^L$. In the absence of steps at the top of the grains, $n_1$ decays by secondary nucleation. Eq. S1 reduces to $\partial_t n_1(t) = \Gamma n'_{step} - D_s n_1^2$, which implies the steady-state value $n_1^s = \sqrt{\Gamma n'_{step}/D_s}$.

The steady-state value of $n_1$ for each scenario is estimated from the following parameters and assumptions. The system is a polycrystalline film of (111)-textured Au (with lattice parameter $a = 0.252$ nm) deposited by physical vapor deposition (PVD) on an incommensurate substrate [e.g. bare Si(100)] at a deposition rate of $F = 1$

ML/min. The PVD takes place at temperatures as low as $T = 100\,°C$ so that thermal re-evaporation is negligible. The diffusion coefficient is estimated to be $D_s \approx 100\,\mu m^2/s$ [Antczak10]. The deposit flux and deposit rate become equivalent under the above conditions. The model further assumes (see details in Figure 1S for a $\xi = 50$ nm-sized grain) the following:

(a) The grains have a round cross-section with stepped ridges forming an annular region around the GB, so that $n'_{step} = \frac{4}{\pi \xi^2} \int_0^{2\pi} \int_0^{\xi/2} [n_{step}(r)\Phi(r)r \cdot \partial r]\partial\theta$, where $n_{step}(r)$ denotes the radial profile of the density of step sites, and Phi-function $\Phi(r)$ takes the value 1 within the region under compression [otherwise $\Phi(r) = 0$]. From Fig. 1S, $n'_{step} \approx 1.3 \times 10^{-3}$ site/lattice$^2$ (33% of $n_{step}$ in the grain).

(b) The strain-energy in the ridges under compression cancels the detachment barrier, so that the detachment rate in these regions is similar to the hopping rate ($\Gamma \approx 4D_s/a^2$).

From these data, we estimate that the flux-supplied steady-state, $n_1^s \approx 5 \times 10^{13}$ adatom/m$^2$, is higher than detachment-supplied steady-state, $n_1^s \approx 1.4 \times 10^{13}$ adatom/m$^2$.

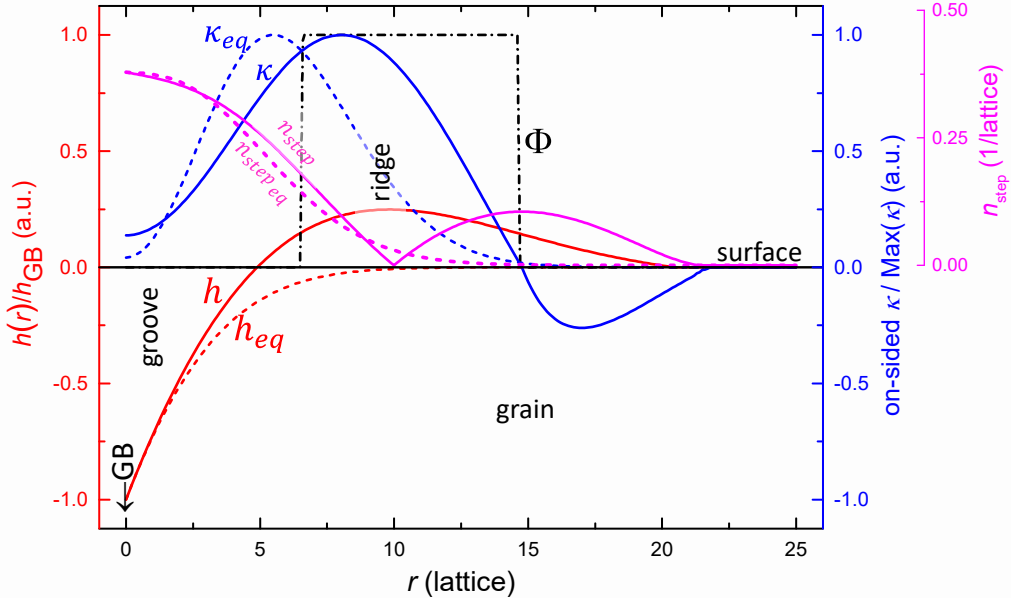

**Figure 1S.** Profiles of height [$h(r)$—red], one-sided curvature [$\kappa(r)$—blue], and density of step sites [$n_{step}(r)$—magenta] on the flank of a $\xi = 50$ nm-sized grain (starting from the triple-junction point at $r = 0$). Solid curves correspond to the kinetic profiles during deposit (without subindex), while dashed ones depict the equilibrium profiles (subscript "eq"). The dashed-dotted black curve corresponds to the Phi-function $\Phi(r)$, which is 1 within the region under compression (where $\kappa > \kappa_{eq}$, according to the Laplace-Young equation), otherwise 0.

**(B) Characteristic time of morphology relaxation in polycrystalline thin films**

The characteristic time $\tau = h/\partial_t h$ of morphology relaxation of the ridges under compression can be estimated from Eqs. 1(a-c) in the manuscript, considering that $\nabla^2 \sigma \approx \sigma/\omega^2$. Thus,

$$\tau(\sigma) = \psi/\sigma \qquad (S2),$$

where $\psi = h_{ridge}\omega^2 k_B T/(D_s n_1^s \Omega^2)$ with $n_1^s = \sqrt{\Gamma n_{step}'/D_s}$ being the steady-state density of adatoms estimated in the previous section for the "after deposit" scenario. $\psi$ can be estimated as $\psi = (n_{ex}/n_1^s)(k_B T/D_s \Omega)$ with $n_{ex} \approx h_{ridge}\omega^2/\Omega$ being computed from height profiles: namely, the $h(r)$—kinetic and $h_{eq}(r)$—equilibrium surface profiles in Fig. 1S. This is done using the relation $n_{ex} = \frac{1}{\Omega}\int_0^{2\pi}\int_0^{\xi/2}\{[h(r) - h_{eq}(r)]r \cdot \partial r\}\partial\theta$. Eq. S2 is supported by Refs. [Chason12, Vasco18, Chason20], and shows that the stress-relaxation kinetics improves (i.e., its rate increases while its characteristic time decreases) as the strength of stress to be released becomes greater.

**(C) Steady-state density of adatoms and characteristic time of relaxation in nanocrystalline thin films**

As discussed in the manuscript, as grain size decreases, the length scales involved in the transports by $\vec{j}_\sigma^L$ and $\vec{j}_k$ towards and from (respectively) different GBs overlap, resulting in an exchange of adatoms between GBs. More stressed GBs relax at the expense of generating new stress in other GBs, so the net stress is homogenized rather than relaxed. Eq. S1 has to be rewritten to account for this adatom exchange so that the second term is no longer null for the "after deposit" scenario. Thus, we write $\partial_t n_1(t) = \Gamma n_{step}' - \epsilon D_s n_1 n_{step} - D_s n_1^2$, where $\epsilon = (\vec{\nabla} \cdot \vec{j}_\sigma^L)/(\vec{\nabla} \cdot \vec{j}_k) \approx (2\lambda/\xi)(J_\sigma^L/J_k)$ is a dimensionless exchange coefficient describing the proportion of adatoms transported by $\vec{j}_\sigma^L$ that diffuse back (by $\vec{j}_k$) towards the less stressed GBs. As a result of the adatom exchange, the steady-state density is $n_1^s = \sqrt{\Gamma n_{step}'/D_s + (\epsilon n_{step}/2)^2} - (\epsilon n_{step}/2)$, which agrees with the results above (in section A) for polycrystalline films (with $\epsilon = 0$). As expected, adatom exchange ($1 \geq \epsilon > 0$) induces a decrease (increase) in $n_1^s$ ($\psi$ in Eq. S2), significantly increasing the characteristic relaxation time. Just to give an example (estimate made from the parameters in section A), the characteristic time for the $\xi =$ 50 nm-sized grain shown in Fig. 1S to release an intrinsic stress of $\sigma = 10$ MPa increases from 174 s for $\epsilon = 0$ (polycrystalline behavior) up to 74486 s (≈20.7 h) for $\epsilon = 0.1$ (nanocrystalline behavior).